\documentclass[
    ,final            
  ]
  {aipproc}

\layoutstyle{8x11single}
\usepackage{amsmath}
\usepackage{lscape,colordvi}

\def\vspaa{\vspace{-.7cm}}

\def\etp{\enlargethispage{.00064cm}}
\usepackage{verbatim}

\begin{document}

\title[Le Delliou et~al.: Separating expansion from contraction \dots]{Separating expansion from contraction: generalized TOV condition,
LTB models with pressure and $\Lambda$CDM}

\classification{98.80.-k,~98.80.Cq,~98.80.Jk,~95.30.Sf~,~04.40.Nr,~04.20.Jb\hfill {\bf Preprint} IFT-UAM/CSIC-09-23\vspaa}

\author{Morgan Le Delliou}{
  address={\textbf{Speaker};Instituto de F\'isica Te\'orica UAM/CSIC, Facultad
de Ciencias, C-XI, Universidad Aut\'onoma de Madrid, Cantoblanco, 28049
Madrid SPAIN. Email: Morgan.LeDelliou@uam.es},
,altaddress={Centro de F\'isica Te\'orica e Computacional, Universidade de Lisboa,  Av. Gama Pinto 2, 1649-003 Lisboa, Portugal.} 
}

\author{Filipe C. Mena}{
  address={Centro de Matem\'{a}tica, Universidade do Minho,
Campus de Gualtar, 4710-057 Braga, Portugal Email: fmena@math.uminho.pt}
}

\author{Jos\'e P. Mimoso}{
  address={Departamento de F\'isica, Faculdade de Ci\^encias, Edif\'icio C8, Campo
Grande, P-1749-016 Lisboa, Portugal Email: jpmimoso@cii.fc.ul.pt},
,altaddress={Centro de F\'isica Te\'orica e Computacional,
Universidade de Lisboa,  Av. Gama Pinto 2, 1649-003 Lisboa,
Portugal.} }


\begin{abstract}
We discuss the existence of a dividing shell separating expanding
and collapsing regions in spherically symmetric solutions with pressure.
We obtain gauge invariant conditions relating not only the intrinsic
spatial curvature of the shells to the ADM mass, but also a function
of the pressure which we introduce that generalises the Tolman-Oppenheimer-Volkoff
equilibrium condition, in the framework of a 3+1 spacetime splitting.
We consider the particular case of a Lemaître-Tolman-Bondi dust models
with a cosmological constant (a $\Lambda$-CDM model) as an example
of our results. 

\end{abstract}

\maketitle

\section{Introduction
}

Cosmological structure formation assumes the collapse of inhomogeneities,
via gravitational instability, into {}``bound'' structures, with
the underlying idea that they depart from the cosmological expansion.
This approach usually models overdensities within closed patches embedded
in Friedman backgrounds, in particular in the spherical collapse included
in the Press \& Schechter scheme \cite{PS74}. Birkhoff's theorem
is often invoked to claim that the evolution of the overdensities
is independent \cite{PeeblesPadm}, while, rigourously, it only applies
to asymptotically flat spacetimes from the cosmic expansion \cite{Birkhoff23}.%
{}

In this work, we define general conditions for the existence of a
shell separating collapse from expansion, and will illustrate our
results with a simple example of inhomogeneous $\Lambda$-CDM models.
Our work differs from previous approaches (see e.g. \cite{Bonnor,Herrera})
since it does not involve spacetime matchings or metric perturbations.
Instead, we adopt the Generalised Painlevé-Gullstrand (hereafter GPG)
formalism used in Lasky \& Lun \cite{LaskyLun06b}, which involves
a $3+1$ splitting (ADM) and the consideration of gauge invariants
kinematic quantities \cite{EllisElst98}. We then define general conditions
for the existence of a shell separating contraction from expansion,
before proposing particular examples of inhomogeneous $\Lambda$-CDM
models implemented as Lemaître-Tolman-Bondi models with dust and a
cosmological constant. Finally, we perform a dynamical study of such
models. 

\section{ADM approach to LTB models in GPG system
}

\etp
\noindent %
\begin{figure}
\begin{centering}
\includegraphics[bb=146bp 342bp 821bp 683bp,clip,width=0.4\paperwidth]
{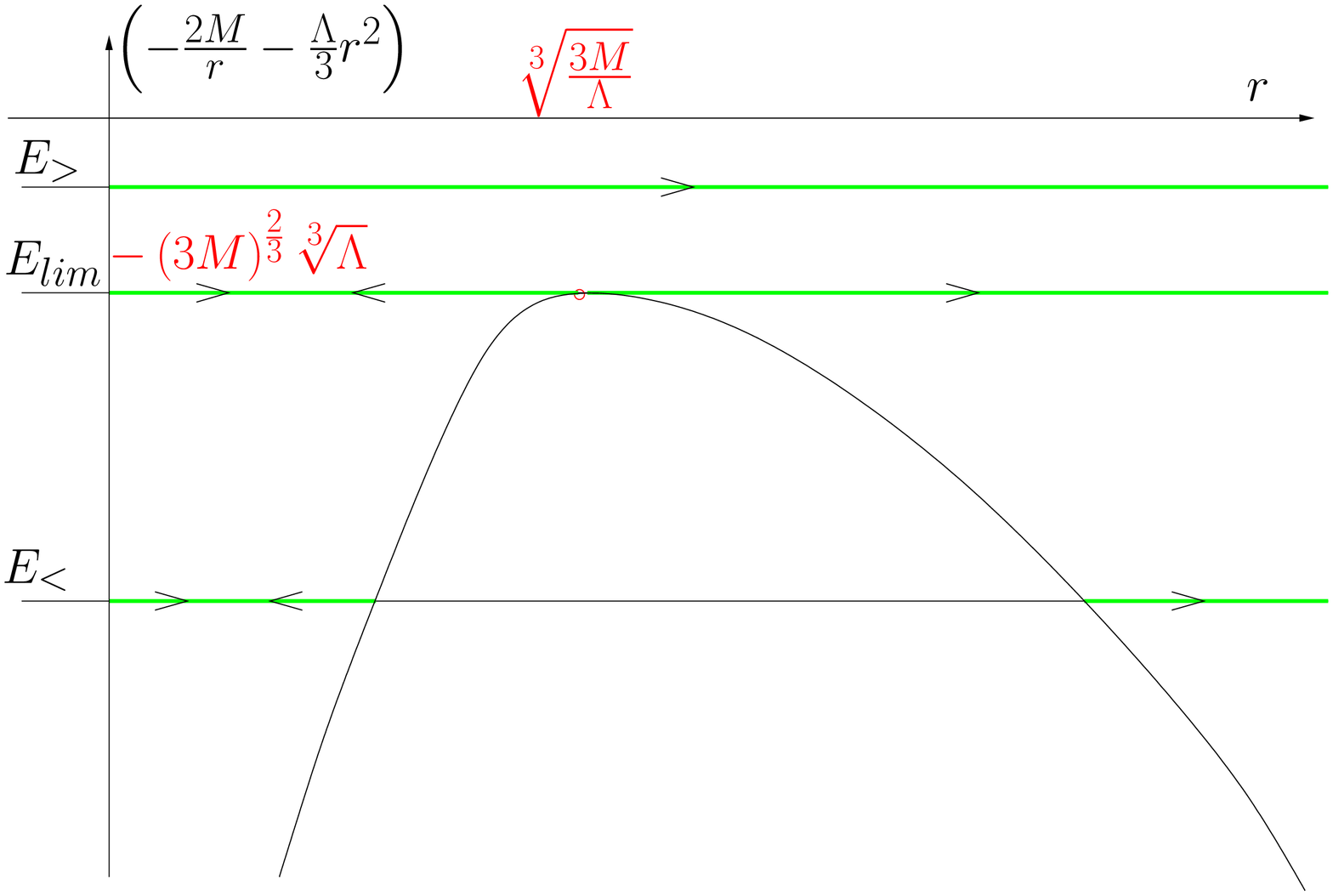}
\par\end{centering}

\caption{\label{fig:KinAn}Kinematic analysis of motion in the
pseudo-potential V. Depending on $E$ relative to $E_{lim}$, for a
given shell of constant $M$ and $E$, the fate of the shell is either
to remain bound ($E_{<}<E_{lim}$) or to escape and cosmologically
expand ($E_{>}>E_{lim}$). There exists a critical behaviour where
the shell will forever expand, but within a finite, bound radius
($E=E_{lim}$, $r\le r_{lim}$).}

\end{figure}
We consider a spherically symmetric Generalised Lemaître-Tolman-Bondi
metric to include pressure. Performing an ADM 3+1 splitting in the
GPG coordinates~\cite{LaskyLun06b} , the metric reads\begin{align}
ds^{2} & =-\alpha(t,r)^{2}dt^{2}+\frac{1}{1+E(t,r)}\left(\beta(t,r)dt+dr\right)^{2}+r^{2}d\Omega^{2}\;,\end{align}

\noindent where $\alpha$ is a lapse function, $\beta$ is a shift,
and $E>-1$ is a curvature-energy characterising the curvature of
the spatial surfaces orthogonal to the direction $n_{a}=(-\alpha,0,0,0)$
of the flow. For a perfect fluid, the projected Bianchi identities
$T_{b;a}^{a}=0$ yield the energy density conservation equation after
projecting along the flow $n^{b}$, and the Euler equation projecting
orthogonally to $n^{b}$. Using the projection $h_{a}^{\, b}$, and
the Lie derivative of the density $\rho$ along the flow $\mathcal{L}_{n}\rho$,
we have: \begin{align}
n^{b}T_{b;a}^{a}= & -\mathcal{L}_{n}\rho-\left(\rho+P\right)\Theta=0, & h_{a}^{\, b}T_{b;c}^{c}= & 0\Rightarrow P^{\prime}=-\left(\rho+P\right)\frac{\alpha^{\prime}}{\alpha}\;,\label{eq:Bianchi}\end{align}

\noindent where $P$ is the pressure, the radial derivatives are denoted
by a prime, $^{\prime}$, the time derivatives are represented by
a dot, $\dot{}$, and the expansion is $\Theta$.

Introducing the ADM (also called Misner-Sharp) mass \begin{align}
M & =r^{2}\left(1+E\right)\left(\ln\alpha\right)^{\prime}-4\pi Pr^{3}+\frac{1}{3}\Lambda r^{3}+r^{2}\mathcal{L}_{n}\left(\frac{\beta}{\alpha}\right)\;,\end{align}

\noindent where $\Lambda>0$ is the cosmological constant, we write
Einstein's field equations (EFEs) as Lie derivatives along the flow
and the radial evolution \begin{align}
\mathcal{L}_{n}{E}= & 2\left(\frac{\beta}{\alpha}\right)\frac{1+E}{\rho+P}P^{\prime}, & \Rightarrow\dot{E}= & \beta\left(E^{\prime}+2\frac{1+E}{\rho+P}P^{\prime}\right),\\
\mathcal{L}_{n}M= & 4\pi Pr^{2}\left(\frac{\beta}{\alpha}\right), & \Rightarrow\dot{M}= & \beta\left(M^{\prime}+4\pi Pr^{2}\right),\\
\left(\frac{\beta}{\alpha}\right)^{2}= & E+2\frac{M}{r}+\frac{1}{3}\Lambda r^{2}.\label{eq:radial}\end{align}

\noindent The system becomes then closed when an equation of state
is supplied. The $\Lambda$ term can be absorbed in $M$, $E$ and
$P$, and many fluids mass equations written with each component using
the $\frac{\beta}{\alpha}$ term for the overall sum of the masses.

\section{Separating collapse from expansion
}
\noindent \etp To characterise the separation of expansion from collapse,
we shall use GLTB coordinates in which the flow direction reduces
to $\partial_{T}$ by choosing $\beta=-\dot{r}$. We thus have \begin{align}
ds^{2} & =-\alpha(T,R)^{2}\left(\partial_{T}t\right)^{2}dT^{2}+\frac{\left(\partial_{R}r\right)^{2}}{1+E(T,R)}dR^{2}+r^{2}d\Omega^{2},\\
\dot{M}= & \beta4\pi Pr^{2},\,\dot{E}r^{\prime}=2\beta\frac{1+E}{\rho+P}P^{\prime},\,\left(-\frac{\dot{r}}{\alpha}\right)^{2}=E+2\frac{M}{r}+\frac{1}{3}\Lambda r^{2}.\label{eq:radialLTB}\end{align}
\begin{figure}
\begin{centering}
\includegraphics[width=0.4\columnwidth]
{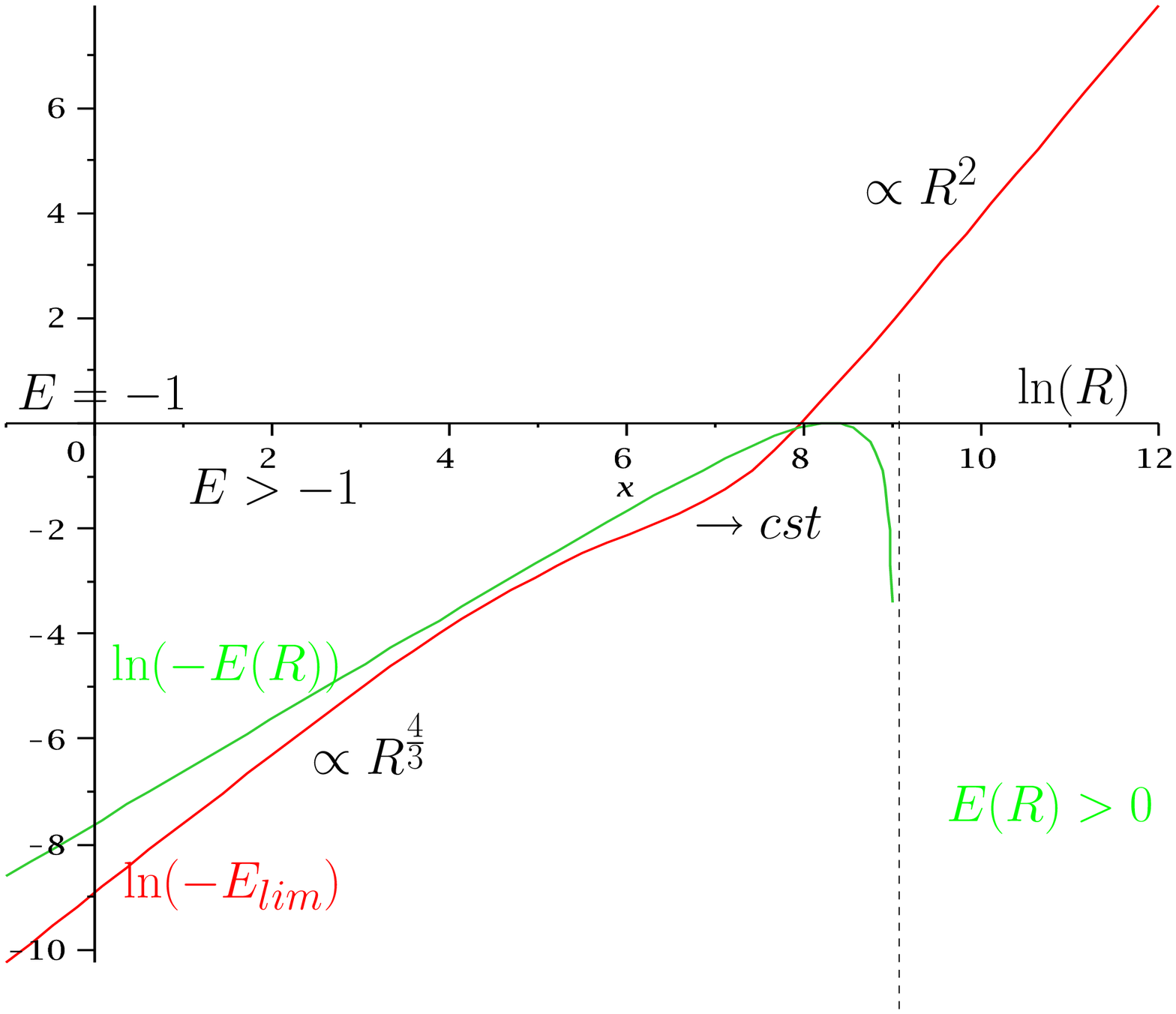}
~\includegraphics[width=0.4\columnwidth]
{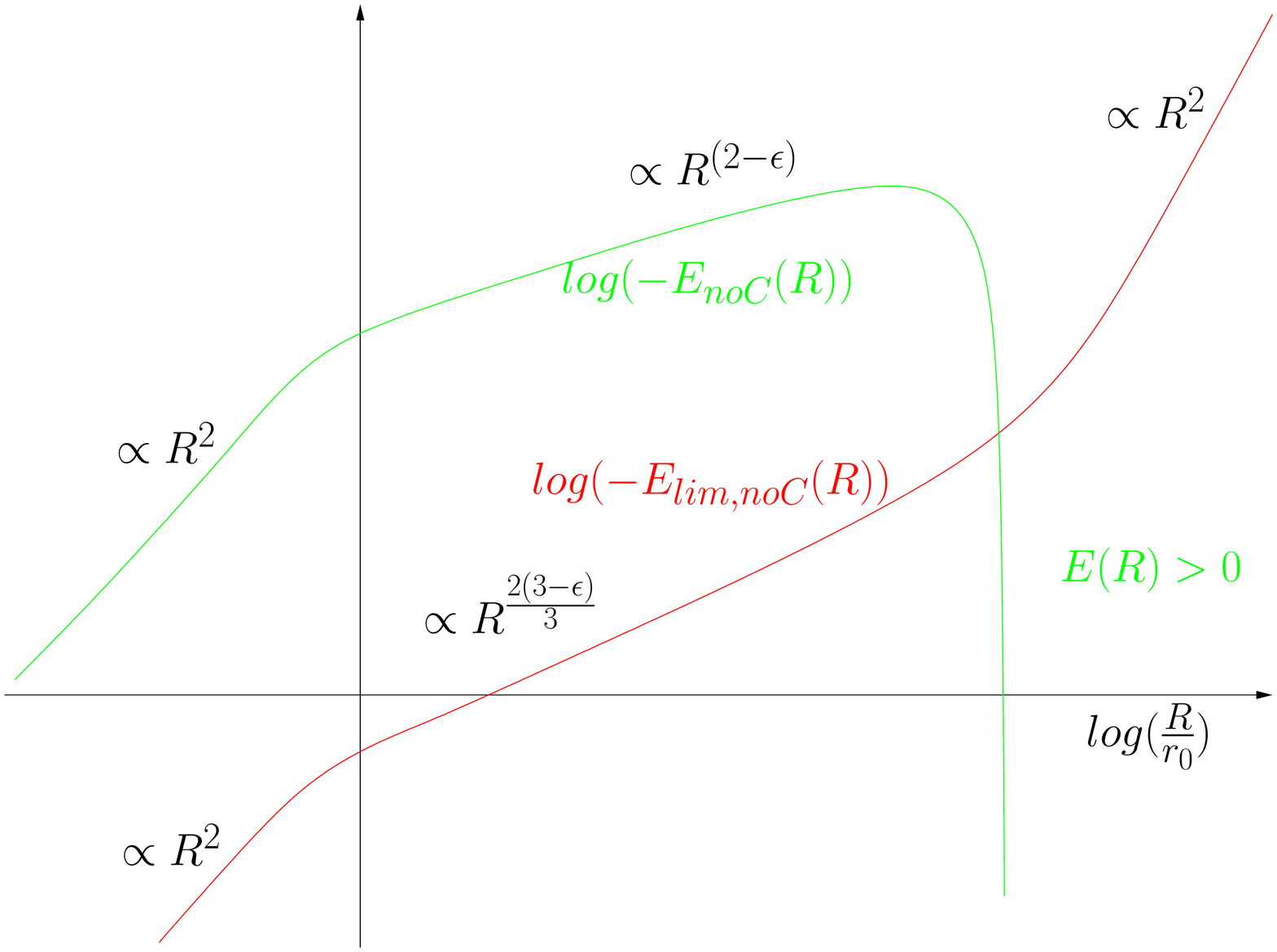}
\par\end{centering}

\caption{\label{cap:NFW+bEmax}\label{cap:NFW+bE+Emax} Plots of
$\log(-E_{lim})-\log(R)$ and $\log(-E)-\log(R)$ for a) NFW \cite{NFW} with background and an $E$ profile given by
$E(R)=-4E_{min}\left(\frac{R}{r_{1}}\right)\left(1-\frac{R}{r_{1}}\right)$,
for $E_{min}=-1+e^{-10}$ and $r_{1}=e^{9}$, and for b)power law
density without cusp + background and Hubble initial flow.}

\end{figure}
where $t$ and $r$ are functions of the coordinates $T$ and $R$.
There are two situations one should consider in parallel. On the one
hand, we look for the gauge invariant expansion $\Theta$, defined
as ${n^{a}}_{;a}$, since our goal is to separate an inner collapsing,
spherical region from the outer expanding universe. On the other hand,
the total ADM mass of the spherical region that departs from the expansion
flow should be conserved. This is indeed suggested by the dust case
where that happens for every shell.

Denoting with $\star$ an evaluation at the dividing shell, we find
the following relation between the expansion $\Theta$ and the shear
$a$, in that at the $\star$ shell\begin{align}
r\left(\frac{\Theta}{3}+a\right) & =-\frac{\beta}{\alpha}=\mathcal{L}_{n}r\Rightarrow & \Theta_{\star}+3a_{\star}=0 & \Leftrightarrow\left.\frac{\beta}{\alpha}\right|_{\star}=0\Leftrightarrow\left.\mathcal{L}_{n}r\right|_{\star}=0\textrm{ when }\mathcal{L}_{n}M(t,r_{\star}(t))=0.\end{align}

\noindent Moreover the generalized Friedman constraint\begin{align}
^{(3)}R+\frac{2}{3}\Theta^{2}= & 6a^{2}+16\pi\rho+2\Lambda\end{align}
tells us that the vanishing of $\Theta$ only happens in regions of
positive 3-curvature $^{(3)}R$. On the other hand, if we demand that
the separating shell has a dust-like vanishing mass/energy flow, i.e.,
has a conserved ADM mass along $n^{a}$: \begin{align}
\forall t,\,\mathcal{L}_{n}M(t,r_{\star}(t))= & 0 & \Rightarrow\forall t,\, E= & -2\frac{M}{r_{\star}}-\frac{1}{3}\Lambda r_{\star}^{2}<0.\label{eq:Estar}\end{align}
We remark that $M$ refers to the total ADM mass, thus including a
cosmological constant. 

%
\begin{figure}
\begin{centering}
\includegraphics[width=0.59\columnwidth]
{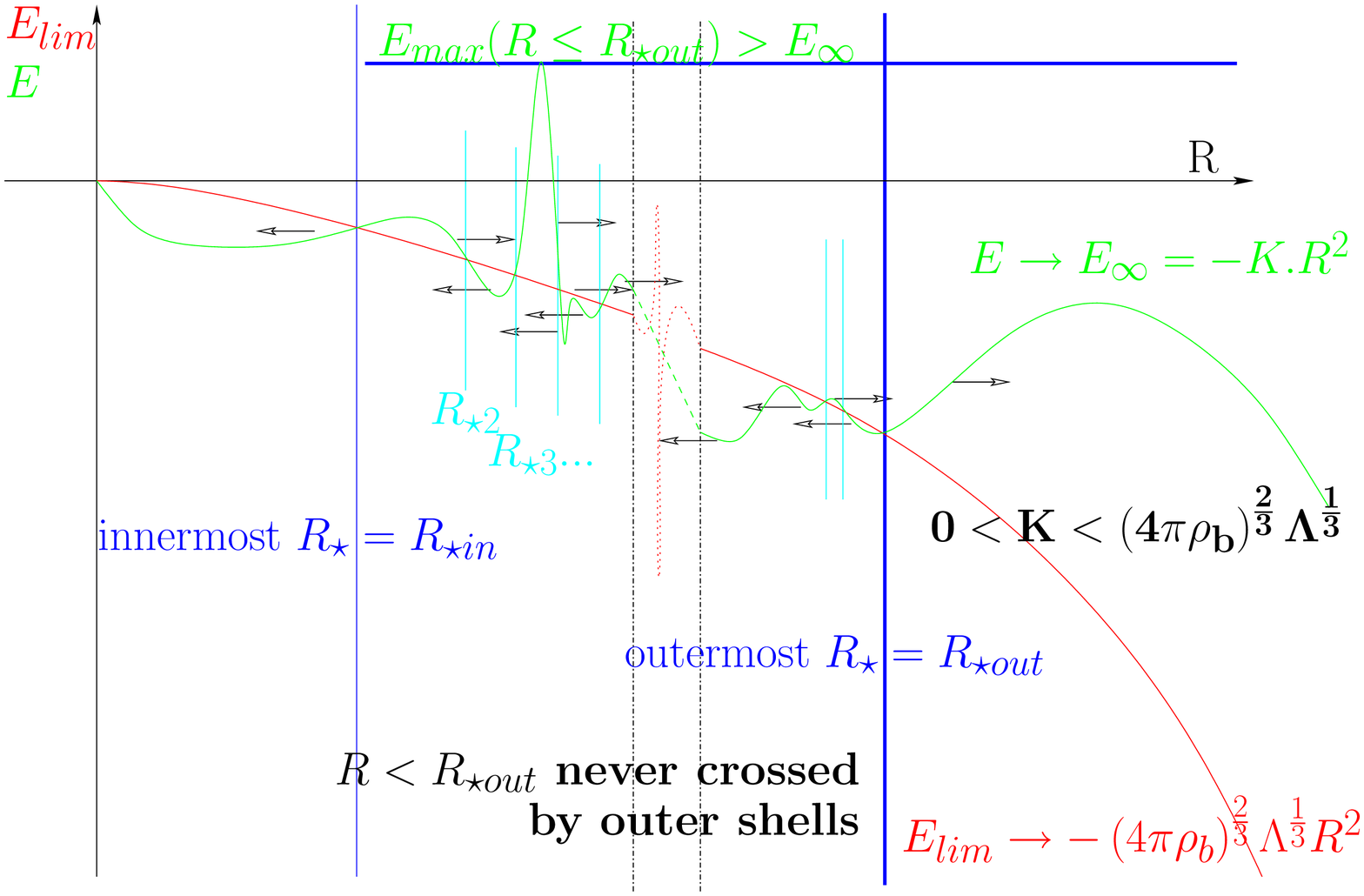}
~\includegraphics[width=0.59\columnwidth]
{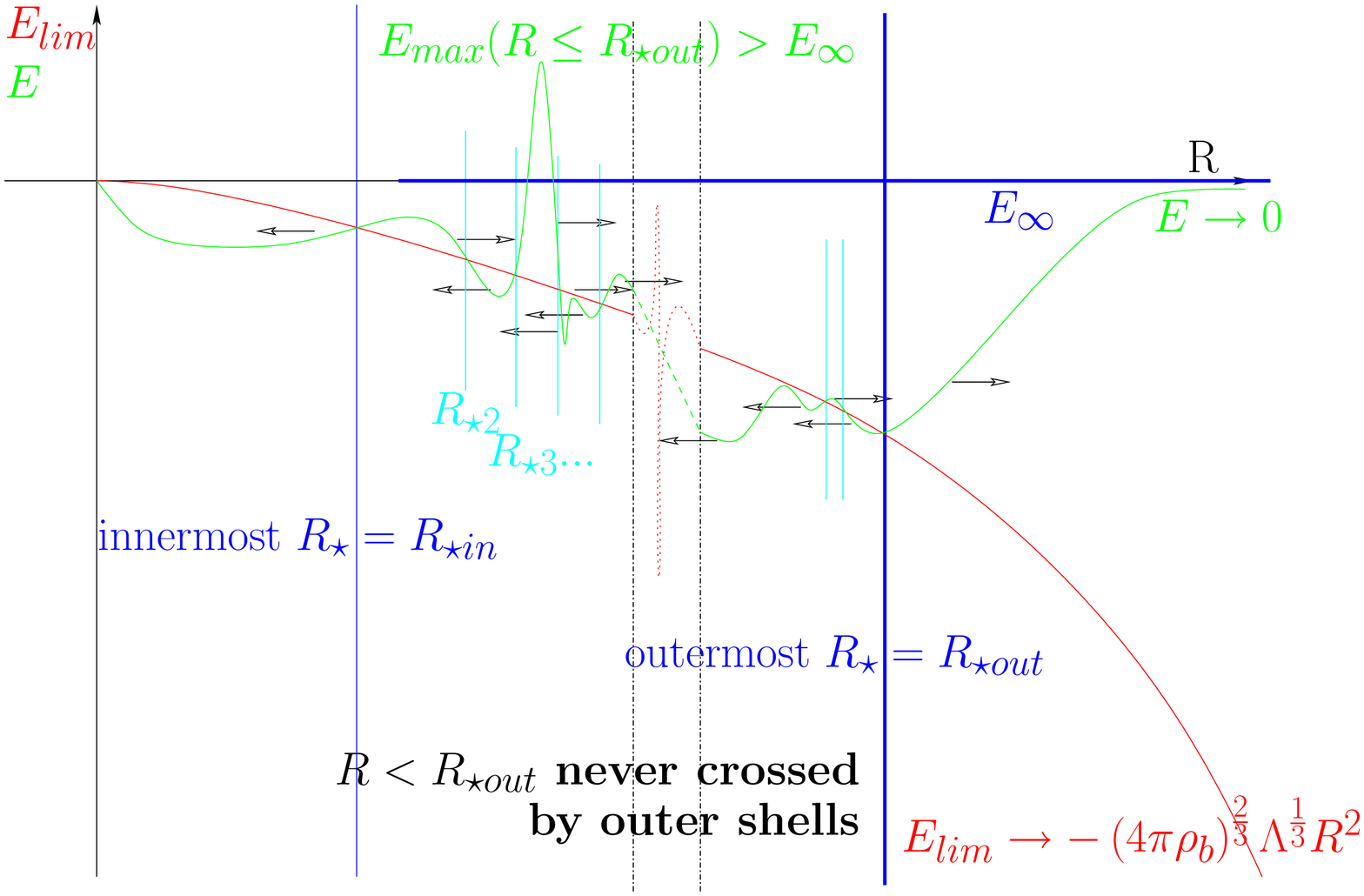}
\par\end{centering}

\caption{%
\label{cap:ClosedOk}Closed and \label{cap:FlatNo}flat backgrounds
with outwards escaping shells: protected inner shells.
}
\end{figure}
Taking into account that the equilibrium of static spherical configurations
requires the satisfaction of the Tolman-Oppenheimer-Volkoff equation
of state \cite{TOV}, we are led to define a generalized gTOV function\begin{align}
\mathrm{gTOV}= & \left[\frac{1+E}{\rho+P}P^{\prime}+4\pi Pr+\frac{M}{r^{2}}-\frac{1}{3}\Lambda r\right]=\mathcal{L}_{n}\left(\frac{\beta}{\alpha}\right)=-\mathcal{L}_{n}^{2}r\;,\label{eq:TOVdef}\end{align}
which reduces to the usual TOV equation when it vanishes on the $\star$-shell.

The radial behaviour of the $\star$-shell is then similar to a turnaround
shell. Indeed $r_{\star}=-\frac{2M_{\star}}{E_{\star}}$ leads to
a null radial velocity $\dot{r}_{\star}=0$ while its acceleration
reveals the importance of the gTOV parameter\begin{align}
\ddot{r}_{\star}= & -\frac{\alpha^{2}}{1+\frac{\Lambda r_{\star}^{2}}{E_{\star}}}\frac{1-\frac{\Lambda}{3M_{\star}}r_{\star}^{3}}{1+\frac{\Lambda}{6M_{\star}}r_{\star}^{3}}\left[\mathrm{gTOV}_{\star}-r_{\star}^{2}\frac{\mathrm{gTOV}_{\star}^{2}}{M_{\star}-\frac{\Lambda}{3}r_{\star}^{3}}\right]; & \ddot{r}_{GLTB,\star}= & -\alpha^{2}\mathrm{gTOV}_{\star}.\end{align}
 The local staticity of the $\star$-shell is then shown
to be equivalent to having a local TOV equation on this limit shell\begin{align}
\mathrm{gTOV}_{\star}=0\Leftrightarrow & \mathcal{L}_{n}^{2}r_{\star}=0\Leftrightarrow\mathcal{L}_{n}\left(\Theta+3a\right)_{\star}=0 & \Leftrightarrow-\frac{1}{\rho+P}P^{\prime}= & \left[\frac{4\pi Pr+\frac{M}{r^{2}}-\frac{1}{3}\Lambda r}{1-\frac{2M}{r}-\frac{1}{3}\Lambda r}\right]_{\star}.\end{align}

\section{An example: $\Lambda$CDM
}
A simple illustration of our result is given by the case of dust with
a $\Lambda$. There is then no pressure gradients and $M_{dust}$
and $E$ are conserved, i.e. $\alpha=1$, which simplifies the analysis
and allow us to perform a kinematic study (see Fig. \ref{fig:KinAn})
per shell of Eq. (\ref{eq:radialLTB}-c). An effective potential is
defined by $V(r)\equiv-\frac{2M}{r}-\frac{\Lambda}{3}r^{2}$, so that
when $\dot{r}=0$ we have indeed $V(r)=E$:\etp\begin{align}
\dot{r}^{2}= & 2\frac{M}{r}+\frac{1}{3}\Lambda r^{2}+E, & \textrm{with }\ddot{r}= & -\frac{M}{r^{2}}+\frac{\Lambda}{3}r.\end{align}
For each shell there is a virtual static state at $r_{lim}=\sqrt[3]{\frac{3M}{\Lambda}}$,
$E_{lim}=-\left(3M\right)^{\frac{2}{3}}\Lambda^{\frac{1}{3}}$, which
only depends on $\Lambda$ and $M(R)$. We also recall that, in this
model, $\mathrm{gTOV}=\frac{M}{r^{2}}-\frac{\Lambda}{3}r=-\ddot{r}$.

\begin{figure}
\begin{centering}
\includegraphics[width=0.59\columnwidth]
{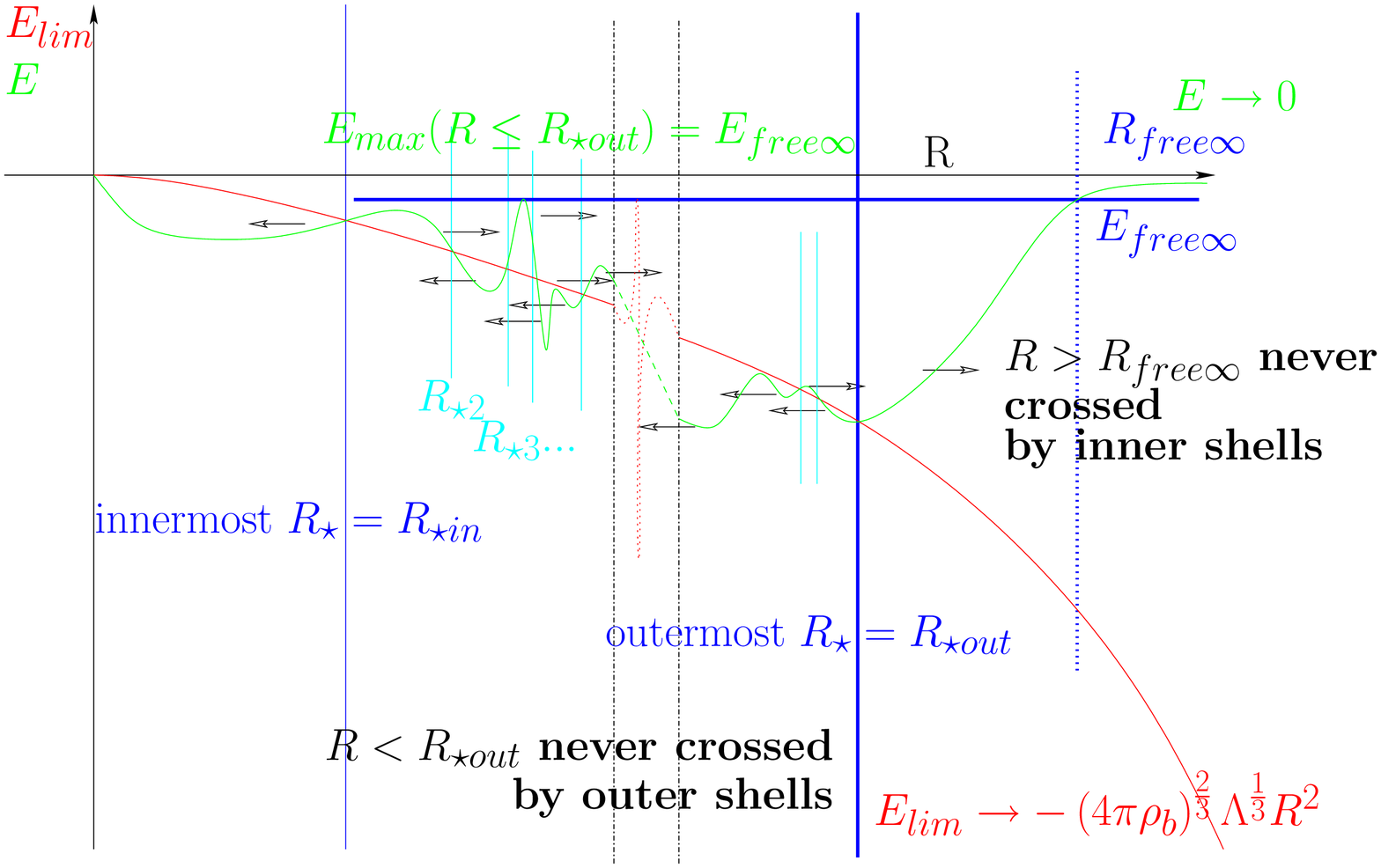}
~\includegraphics[width=0.59\columnwidth]
{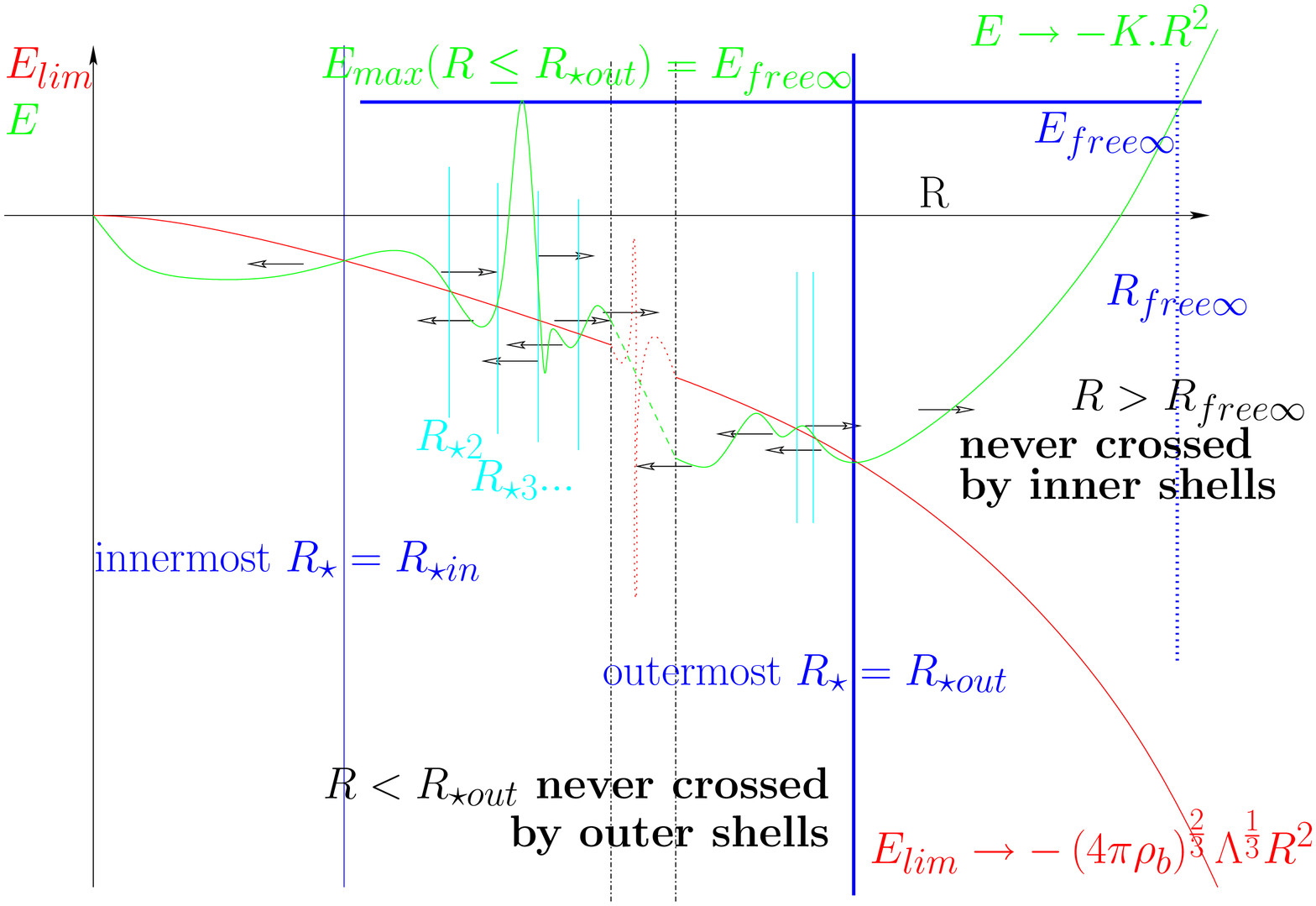}
\par\end{centering}

\caption{%
\label{cap:FlatOk}Flat, inner shells free, and \label{cap:Open}open
backgrounds: two limit shells
}
\end{figure}

We then need to choose initial conditions $\rho_{i}$, which set the
$E_{lim}$ profile, and $v_{i}$, which set the $E_{i}$ profile.
The intersection of $E_{lim}$ with $E_{i}$ will be static at turnaround,
hence defining our limit shell. We use two sets of cosmologically
motivated initial conditions: one with $\rho_{i}$ as a well known
{}``universal'' CDM halo density profile from simulations, refered
to as NFW profile \cite{NFW} and $E_{i}$ as a parabola (Fig. \ref{cap:NFW+bEmax}-a),
and the second, with a cuspless power law as $\rho_{i}$ and a Hubble
flow for $v_{i}$ (Fig. \ref{cap:NFW+bE+Emax}-b). Explorations of
all general cases for initial conditions yields a split between the
separation of inner shells from outer influence, which is guaranteed
for ever expanding backgrounds (thus unless the background recollapses,
see Figs. \ref{cap:ClosedOk} and \ref{cap:FlatOk}) and that of outer
shells from inner influence, which occur for open and some flat backgrounds
(see Figs. \ref{cap:FlatOk}).


\section{Conclusions
}


%
Using non-singular, generalized Painlevé-Gullstrand coordinate formulation
of the ADM spherically symmetric, perfect fluid system \cite{LaskyLun06b}
we have shown \cite{MimLeD,Delliou:2009dm} that the existence of
shells locally separating between inner collapsing and outer expanding
regions, is governed by the condition that the combination of expansion
scalar and shear $\theta+3a$ should vanish on the shell. The ADM
mass of the shell is then conserved. This condition requires that
the separating shell must be located in an elliptic ($E<0$) region.
Moreover, for that shell to exist over time, we have shown that the
TOV equation must be locally satisfied. 
We argue in some cases that this local condition is global in a cosmological
context (FLRW match at radial asymptote). Given appropriate initial
conditions, this translates into global separations between an expanding
outer region and an eventually collapsing inner region. We present
simple but physically interesting illustrations of the results, a
model of Lemaître-Tolman-Bondi dust with $\Lambda$ representing spherical
perturbations in a $\Lambda$CDM model with two different initial
sets of cosmologically interesting conditions consistent with known
phenomenological constraints \cite[and Refs. therein]{MimLeD,Delliou:2009dm}:
an NFW density profile with a simple curvature profile going from
bound to unbound conditions, and a non cuspy power law fluctuation
with initial Hubble flow. We show, for these models, the existence
of a global separation. We argue that these shells are trapped matter
surfaces \cite{MimLeD,Delliou:2009dm} and that, therefore they separate
domains of influence of cosmic expansion from local domains of matter
dynamics and can hold the place, in structure formation studies, of
the incorrect invocation of Birkhoff's theorem. 


\begin{theacknowledgments}


The work of MLeD is supported by CSIC (Spain) under the contract
JAEDoc072, with partial support from CICYT project FPA2006-05807, at
the IFT, Universidad Autonoma de Madrid, Spain, and was also
supported by FCT (Portugal) under the grant SFRH/BD/16630/2004, at
the CFTC, Lisbon University, Portugal. Financial support from the
Foundation of the University of Lisbon and the Portuguese Foundation
for Science and Technology (FCT) under contract POCTI/ISFL/2/618 is also gratefully acknowledged.
FCM is supported by CMAT, Univ. Minho, and FCT project
PTDC/MAT/108921/2008.

\end{theacknowledgments}





\begin{thebibliography}{10}
\bibitem{PS74}W.~H.~Press and P.~Schechter, 
 \emph{ApJ.} \textbf{187},  425 (1974). 


\bibitem{PeeblesPadm}P.~J.~E.~Peebles, \emph{The large-scale structure
of the universe}, Princeton U. Press, Princeton, N.J., 1980; T.~Padmanabhan,\emph{
Structure Formation in the Universe}, Cambridge U. Press, Cambridge,
UK, 1993.

\bibitem{Birkhoff23}G.~D.~Birkhoff, \emph{Relativity and Modern
Physics}, Harvard U. Press, Cambridge, MA, 1923.


%
{}
\bibitem{Bonnor} 
W. B. Bonnor, Mon. Not. R. Astr. Soc. 282 (1996) 1467-1469.

\bibitem{Herrera} A. Abreu, H. Hernandez \& L. A. Nunez, {\em Class. Quant.
Grav.} \textbf{24} (2007) 4631-4645 ; Di Prisco, L. Herrera \& V.
Varela, {\em Gen. Rel. Grav.}, \textbf{29} (1997) 1239

\bibitem{LaskyLun06b}P.~D.~Lasky and A.~W.~C.~Lun, 2006, \emph{PRD}\textbf{
74}, 084013 (2006).


\bibitem{EllisElst98} G.~F.~R.~Ellis and H.~van Elst, 
 NATO 
 Adv.\ Study I.S.C.M.P.S.\ \textbf{541} (1999) 1 {[}arXiv:gr-qc/9812046].

\bibitem{NFW}J.~F.~Navarro, C.~S.~Frenk and S.~D.~M.~White,
\emph{ApJ} \textbf{462}, 563 (1996).


\bibitem{TOV}R.~C.~Tolman, 
 \emph{PR} \textbf{55}, 364 (1939). 
; J.~R.~Oppenheimer and G.~M.~Volkoff, 
 \emph{PR} \textbf{55}, 374 (1939). 

\bibitem{MimLeD}J.~P.~Mimoso, M.~Le Delliou and F.~C.~Mena, Submitted to PRD, 2009 [arXiv:0910.5755]

\bibitem{Delliou:2009dm} M.~L.~Delliou and J.~P.~Mimoso, 
 AIP Conf.\ Proc.\ \textbf{1122}, 316 (2009) {[}arXiv:0903.4651
{[}gr-qc{]}{]}. 

\end{thebibliography}

\end{document}